\documentclass[10pt,twocolumn]{article}
\usepackage[T1]{fontenc}
\usepackage{lmodern}
\usepackage[letterpaper,top=0.70in,bottom=0.74in,left=0.70in,right=0.70in]{geometry}
\usepackage{microtype,amsmath,amssymb,booktabs,array,tabularx,enumitem}
\usepackage{xcolor,hyperref,url,tikz,pgfplots}
\usetikzlibrary{positioning,arrows.meta}
\pgfplotsset{compat=1.18}
\definecolor{StudyBlue}{HTML}{245DA8}
\definecolor{StudyTeal}{HTML}{147B70}
\definecolor{StudyMuted}{HTML}{5E6A75}
\definecolor{StudyGrid}{HTML}{D5DCE3}
\hypersetup{pdftitle={Commercial Intent in Human--AI Conversations: A Corpus Audit and Architecture for Website Sales Agents},pdfauthor={Benjamin Tannenbaum},colorlinks=true,linkcolor=StudyBlue,citecolor=StudyBlue,urlcolor=StudyBlue}
\setlist{nosep,leftmargin=*}

\newcolumntype{Y}{>{\raggedright\arraybackslash}X}
\title{\vspace{-0.45in}\textbf{Commercial Intent in Human--AI Conversations:}\\A Corpus Audit and Architecture for Website Sales Agents}
\author{Benjamin Tannenbaum\\Javo\\\texttt{getjavo.com}}
\date{\vspace{-0.12in}}
\begin{document}
\maketitle
\begin{abstract}
Conversational sales agents must distinguish questions about products from
purchase commitments, preserve explicit requirements, and ground the next
action in current business information. We report an aggregate census of
725,219 records in an accessible conversation table provided by Aiso and
develop a reference architecture for this setting. All records have distinct
non-null conversation hashes. Existing metadata labels identify 41,800
commercial records (5.76\%) and 2,387 transactional records (0.33\%); their
union contains 44,187 records (6.09\%). Within the commercial category,
54.31\% are labeled English, 41.59\% have recorded depth of at least two,
and 13.51\% have depth of at least four. Commercial-label prevalence varies
from 4.87\% to 6.35\% across three source batches. These measurements
motivate explicit separation of corpus inventory, commercial relevance,
training eligibility, and observed business outcomes. The proposed architecture
combines business-grounded knowledge, provenance-bearing conversation state,
and a constrained next-action policy. A quality specification addresses source
rights, privacy, label validation, deduplication, and training-test separation.
The study is a metadata audit and technical design, not a validation of label
accuracy, model training volume, or sales conversion. No raw conversation text
or personal identifiers are released.
\end{abstract}

\section{Introduction}
A website visitor can ask about a price, add a location or budget, reject an
option, and request a call without stating a complete buying brief in one
message. A useful conversational sales agent must interpret those updates,
answer from current business facts, and identify an appropriate next step.
The relevant behavior is therefore broader than fluent response generation.
It includes uncertainty handling, requirement retention, clarification, and
human handoff.

Conversation data can inform these behaviors, but the relationship between
data quantity and commercial capability needs careful accounting. A corpus
record may contain a commercial question without a purchase, a transaction
request without execution, or an assistant answer that is unsuitable for
training. Three claims should remain distinct: the size of a stored corpus,
the number of commercially relevant examples, and the number of examples
actually used to train a particular model version. None of these counts alone
establishes conversion improvement.

This technical white paper pairs original aggregate analysis with a reference
architecture for website sales agents. The empirical contribution is a census
of the accessible main conversation table, with exact denominators, source-batch
comparisons, and metadata-quality checks. The design contribution specifies
how conversation data, business knowledge, state tracking, and action policy
can be separated in a commercial agent.

Javo is the motivating product. Its public description includes answering
website questions, qualifying requests, and collecting lead context
\cite{javo}. Its product owner reports that Javo's team trains its model using
conversation data. This paper does not inspect model weights, training jobs,
training manifests, or deployed architecture. The reference architecture and
quality pipeline are specifications, rather than verified descriptions of the
current implementation. This boundary preserves the usefulness of the design
without converting a product statement into an experimental result.

\section{Related Work}
Research on e-commerce pre-sales dialogue studies the combination of large
language models and conversational recommender systems. Liu et al.
\cite{liu2023} investigate complementary roles for recommendation and language
generation. That work motivates separating a conversation model from
domain-specific evidence, but it does not establish that a general commercial
intent label identifies a successful sale.

SalesLLM \cite{su2026} evaluates multi-turn sales behavior through curated
scenarios, a customer simulator, and scores for sales progress and buying
intent. Its revised version reports human calibration and distinguishes
typical human salespeople from experts. It illustrates the value of evaluating
process and outcome separately. Our audit neither runs this benchmark nor
imports its reported results as evidence for Javo. Simulated buying intent,
human preference, and a completed purchase remain different endpoints.

The contribution here is narrower: explicit accounting of a selected
conversation corpus and a reference architecture tied to those measurement
boundaries. It is not a claim of a new best-performing sales model.

Earlier work by the author distinguishes recorded recommendation content
from observable buyer responses \cite{tannenbaum2026buyer} and experimentally
compares full within-conversation context with isolated final prompts
\cite{tannenbaum2026context}. These studies motivate outcome separation and
context retention, respectively. Their samples and results are not treated
as new observations in the present metadata census, and overlap with their
underlying records was not audited.

\subsection{Conversation, retrieval, and adaptation}
Conversational recommendation treats preference elicitation, explanations,
and feedback as parts of an interaction rather than a single ranking
\cite{jannach2020,jannach2022challenge}. Its evaluation requires attention
to both component quality and users' experience of the complete system
\cite{jannach2022eval}. This distinction is directly relevant to a website
agent: producing a plausible recommendation and identifying an appropriate
business action are separate responsibilities.

Retrieval-augmented generation combines a parametric language model with
external retrieved evidence \cite{lewis2020}. This provides a methodological
basis for separating replaceable business facts from learned conversational
behavior. ReAct studies interleaved reasoning and actions \cite{yao2022},
while Toolformer investigates learning when to invoke tools \cite{schick2023}.
These works motivate explicit interfaces between language generation and
external operations; they do not establish that an unconstrained action
policy is appropriate for sales. Our design therefore places business
permissions and evidence checks outside unrestricted response generation.

DialoGPT studies conversational response generation from large-scale dialogue
\cite{zhang2019}. Instruction tuning with human feedback \cite{ouyang2022}
and chat-model adaptation \cite{touvron2023} illustrate that training data,
behavioral objectives, and evaluation need separate reporting. LoRA offers
an approach to parameter-efficient adaptation \cite{hu2021}; direct preference
optimization provides a method for learning from response preferences
\cite{rafailov2023}. These are possible methods for an implementation, not
methods verified as used by Javo in this study.

\subsection{Documentation and evaluation}
Datasheets for Datasets \cite{gebru2018} motivate documenting collection,
composition, intended uses, and limitations. Model Cards \cite{mitchell2018}
address the corresponding reporting of model use and performance. We adopt
this separation: the present corpus audit documents an accessible inventory,
whereas a model report would require a specific checkpoint and evaluation.
Deduplication research \cite{lee2021} motivates checks beyond unique record
identifiers. Long-context evaluation \cite{liu2023middle} shows why simply
including more history is not sufficient evidence that a model uses relevant
requirements reliably. Holistic evaluation \cite{liang2022} motivates a
set of distinct metrics rather than one aggregate success score. Together,
these sources support our quality specification without supplying missing
empirical validation for this corpus or product.

\section{Data and Audit Scope}
\subsection{Accessible corpus}
On 5 October 2026, read-only database queries enumerated the accessible main
conversation table provided by Aiso. The audit covers 725,219 records from
three stored source batches, anonymized as A, B, and C in public outputs.
Batch sizes are 233,619, 234,553, and 257,047. These are administrative source
groups, not randomized experimental conditions or independent replications.
The table is an accessible inventory scope; it is not a verified census of
every conversation held by the provider or a released training subset.

The audited schema includes a conversation hash, stored intent, language,
source batch, recorded turn count, and a safety flag. Only aggregate counts
were exported. Raw conversation text, account and participant identifiers,
client names, and internal collection names are excluded from the artifacts.
No textual examples from private records appear in the paper.

\subsection{Provenance and disclosure}
The corpus is privately maintained. That description does not imply exclusive
ownership of all underlying text: third-party source material retains its
applicable rights and terms. Source identities, collection histories, and
source-specific research and training permissions belong in a restricted
provenance register. Such a register was not inspected in this audit.
Public reporting here is limited to aggregate metadata. No claim is made
that every raw record is cleared for model training or redistribution.

The existing metadata labels are inherited, not newly annotated for this
study. The audit did not inspect the intent classifier, its version, training
data, or annotation agreement. Missing intent is treated as an unlabeled
category. Recorded depth is the stored turn-count value; its exact role and
message-count semantics were not established. A threshold of two is therefore
not described as proof of two buyer turns.

\subsection{Accounting checks}
A separate global query counted distinct conversation hashes. Both row count
and distinct non-null hash count equal 725,219. This establishes identifier
uniqueness in the audited table, not semantic uniqueness. Translations,
rephrased copies, and overlapping sessions can have different hashes.
Source-level distinct-hash totals reconcile with the global total.

Recorded depth and the safety field are populated in all audited records.
The safety flag is false for every record. This uniformity does not establish
universal safety or privacy: it may reflect upstream filtering, field
construction, or source selection, none of which was independently verified.
A separate stored summary table differed from the underlying-table totals;
the underlying table supplies every number reported here.

\section{Measures and Estimation}
Let $N$ be the number of audited records and $I_i$ the stored intent label of
record $i$. The primary commercial indicator is
\begin{equation}
 C_i=\mathbb{1}(I_i=\texttt{commercial}).
\end{equation}
The transactional indicator $T_i$ is defined analogously. A broader relevance
sensitivity is $B_i=\max(C_i,T_i)$. Because each record has one stored label,
the two categories are disjoint. Their shares are
\begin{equation}
 \widehat{p}_C=\frac{\sum_i C_i}{N},\qquad
 \widehat{p}_B=\frac{\sum_i B_i}{N}.
\end{equation}
These indicators operationalize category membership, not latent buying
intent or a realized transaction.

For category $k$, we report the proportion with an English label and the
proportions whose recorded depth is at least two and four. The denominator
for each conditional share is the number of records labeled $k$. We also
report commercial-label prevalence separately by source batch.

All estimates are descriptive ratios of an enumerated table. We do not
report sampling confidence intervals for this census. Such intervals would
not address unknown collection selection, semantic dependence, or label
error. We do not infer intersections from marginal counts, and do not
interpret batch differences causally. No transcript coding, model generation,
user simulation, or sales-outcome linkage was performed.

\section{Results}
\subsection{Commercial-label composition}
Table~\ref{tab:intents} reconciles all categories with the corpus denominator.
The informational category is largest, followed by other. Commercial records
account for 41,800 of 725,219 records (5.76\%). Transactional records account
for 2,387 (0.33\%). Including both yields 44,187 (6.09\%), compared with
the narrow commercial definition. This sensitivity changes the numerator by
2,387 and the share by approximately 0.33 percentage points.

\begin{table}[t]
\centering\small
\begin{tabular}{lrr}
\toprule Stored intent & Records & Share (\%)\\\midrule
Informational & 376,889 & 51.97\\
Other & 204,774 & 28.24\\
Productive & 55,489 & 7.65\\
Creative & 42,106 & 5.81\\
Commercial & 41,800 & 5.76\\
Transactional & 2,387 & 0.33\\
Navigational & 1,764 & 0.24\\
Unlabeled & 10 & 0.0014\\\midrule
Total & 725,219 & 100.00\\\bottomrule
\end{tabular}
\caption{Intent composition of the audited table. Percentages are rounded
independently. Stored labels are not independently validated ground truth.}
\label{tab:intents}
\end{table}

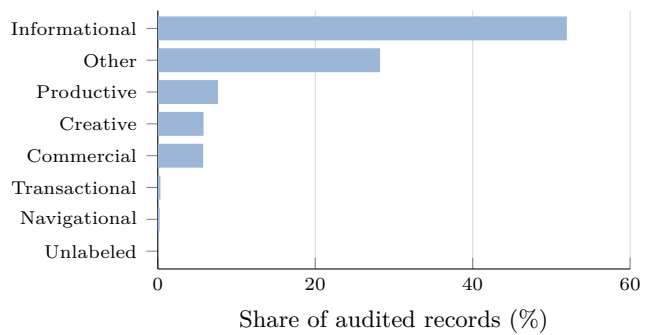
\begin{figure}[t]
\centering
\begin{tikzpicture}
\begin{axis}[width=0.90\columnwidth,height=5.0cm,xbar,bar width=9pt,
 xmin=0,xmax=60,xtick={0,20,40,60},xlabel={Share of audited records (\%)},
 symbolic y coords={Unlabeled,Navigational,Transactional,Commercial,Creative,Productive,Other,Informational},
 ytick=data,tick label style={font=\scriptsize},label style={font=\small},
 xmajorgrids=true,grid style={StudyGrid},axis lines*=left,enlarge y limits=.08]
\addplot[fill=StudyBlue!45,draw=none] coordinates
{(0.0014,Unlabeled)(0.24,Navigational)(0.33,Transactional)(5.76,Commercial)(5.81,Creative)(7.65,Productive)(28.24,Other)(51.97,Informational)};
\end{axis}
\end{tikzpicture}
\caption{Stored intent distribution, $N=725{,}219$. Table~\ref{tab:intents}
gives exact category counts.}
\label{fig:mix}
\end{figure}

\subsection{Language coverage and recorded depth}
The commercial category contains 22,702 English-labeled records (54.31\%).
It contains 17,384 records with depth at least two (41.59\%) and 5,649 with
depth at least four (13.51\%). These are marginal counts; they do not
establish the size of an English, deep-conversation intersection.

Table~\ref{tab:coverage} compares selected categories. Commercial records are
more often labeled English than informational records (54.31\% versus
45.33\%), but are less often at depth at least two (41.59\% versus
43.69\%). Neither difference identifies why a conversation is long, whether
it contains objections, or whether a recommendation succeeded.

\begin{table}[t]
\centering\small
\begin{tabular}{lrrr}
\toprule Intent & English & Depth $\geq2$ & Depth $\geq4$\\\midrule
Commercial & 54.31 & 41.59 & 13.51\\
Transactional & 53.83 & 43.49 & 12.78\\
Informational & 45.33 & 43.69 & 14.88\\
Other & 45.97 & 40.67 & 15.90\\\bottomrule
\end{tabular}
\caption{Conditional shares (\%) within each stored intent. Depth semantics
are not role-validated. Denominators appear in Table~\ref{tab:intents}.}
\label{tab:coverage}
\end{table}

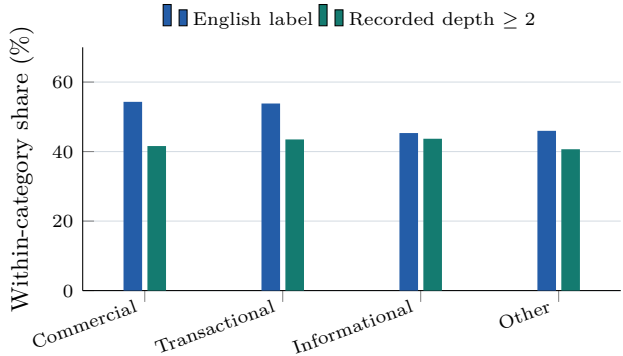
\begin{figure}[t]
\centering
\begin{tikzpicture}
\begin{axis}[width=\columnwidth,height=4.8cm,ybar,bar width=7pt,
 ymin=0,ymax=70,ytick={0,20,40,60},ylabel={Within-category share (\%)},
 symbolic x coords={Commercial,Transactional,Informational,Other},xtick=data,
 x tick label style={font=\scriptsize,rotate=20,anchor=east},
 y tick label style={font=\scriptsize},label style={font=\small},
 ymajorgrids=true,grid style={StudyGrid},axis lines*=left,
 legend style={font=\scriptsize,at={(.5,1.03)},anchor=south,legend columns=2,draw=none},
 enlarge x limits=.15]
\addplot[fill=StudyBlue,draw=none] coordinates {(Commercial,54.31)(Transactional,53.83)(Informational,45.33)(Other,45.97)};
\addplot[fill=StudyTeal,draw=none] coordinates {(Commercial,41.59)(Transactional,43.49)(Informational,43.69)(Other,40.67)};
\legend{English label,Recorded depth $\geq2$}
\end{axis}
\end{tikzpicture}
\caption{Language coverage and recorded depth by intent. Each bar uses
the full size of its own category as denominator.}
\label{fig:coverage}
\end{figure}

\subsection{Source-batch heterogeneity}
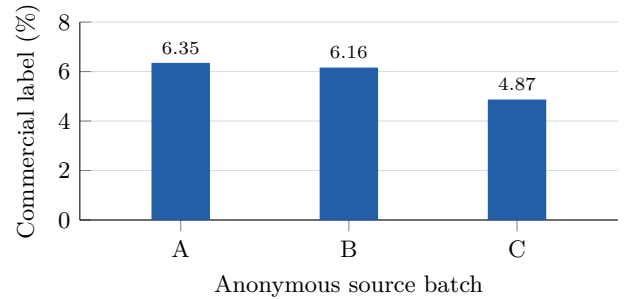
\begin{figure}[t]
\centering
\begin{tikzpicture}
\begin{axis}[width=\columnwidth,height=4.2cm,ybar,bar width=22pt,
 ymin=0,ymax=8,ytick={0,2,4,6,8},ylabel={Commercial label (\%)},
 symbolic x coords={A,B,C},xtick=data,xlabel={Anonymous source batch},
 tick label style={font=\small},label style={font=\small},
 nodes near coords,point meta=explicit symbolic,
 every node near coord/.append style={font=\scriptsize},
 ymajorgrids=true,grid style={StudyGrid},axis lines*=left,enlarge x limits=.3]
\addplot[fill=StudyBlue,draw=none] coordinates {(A,6.35)[6.35] (B,6.16)[6.16] (C,4.87)[4.87]};
\end{axis}
\end{tikzpicture}
\caption{Source-batch commercial-label prevalence. Batch counts are
233,619, 234,553, and 257,047 for A, B, and C, respectively. These are
source groups, not randomized treatments.}
\label{fig:batch}
\end{figure}

Batch A contains 14,837 commercially labeled records out of 233,619 (6.35\%);
batch B contains 14,457 out of 234,553 (6.16\%); and batch C contains
12,506 out of 257,047 (4.87\%). The range is approximately 1.48 percentage
points. Figure~\ref{fig:batch} makes the differing baselines visible.
Source populations, collection design, or annotation differences could
contribute to this variation; the audit does not separate those mechanisms.
The pooled rate should not be interpreted as the prevalence of commercial
intent among all website visitors or all users of conversational AI.

\subsection{Exclusive depth bands and language labels}
\begin{figure}[t]
\centering
\begin{tikzpicture}
\begin{axis}[width=.90\columnwidth,height=4.0cm,xbar,bar width=14pt,
 xmin=0,xmax=70,xtick={0,20,40,60},xlabel={Commercial records (\%)},
 symbolic y coords={At least 4,2 through 3,Below 2},ytick=data,
 tick label style={font=\scriptsize},label style={font=\small},
 nodes near coords,every node near coord/.append style={font=\scriptsize},
 xmajorgrids=true,grid style={StudyGrid},axis lines*=left,enlarge y limits=.3]
\addplot[fill=StudyTeal,draw=none] coordinates {(13.51,At least 4)(28.07,2 through 3)(58.41,Below 2)};
\end{axis}
\end{tikzpicture}
\caption{Exclusive recorded-depth bands among 41,800 commercial records.
Counts are 5,649, 11,735, and 24,416, respectively.}
\label{fig:bands}
\end{figure}
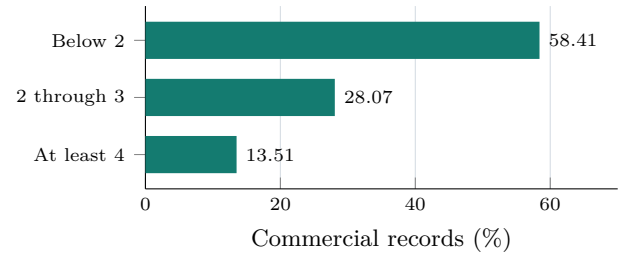

The nested depth thresholds permit an exact partition of the commercial
category. Of its 41,800 records, 24,416 (58.41\%) have recorded depth below
two, 11,735 (28.07\%) have depth from two through three, and 5,649
(13.51\%) have depth of at least four (Figure~\ref{fig:bands}). The middle
band is obtained by subtracting the depth-at-least-four count from the
depth-at-least-two count. These bands are mutually exclusive and exhaustive.
They describe stored depth, not verified stages in a sales funnel.

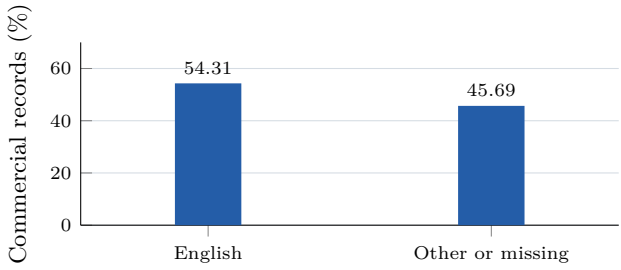
\begin{figure}[t]
\centering
\begin{tikzpicture}
\begin{axis}[width=\columnwidth,height=4.0cm,ybar,bar width=25pt,
 ymin=0,ymax=70,ytick={0,20,40,60},ylabel={Commercial records (\%)},
 symbolic x coords={English,Other or missing},xtick=data,
 tick label style={font=\scriptsize},label style={font=\small},
 nodes near coords,every node near coord/.append style={font=\scriptsize},
 ymajorgrids=true,grid style={StudyGrid},axis lines*=left,enlarge x limits=.45]
\addplot[fill=StudyBlue,draw=none] coordinates {(English,54.31)(Other or missing,45.69)};
\end{axis}
\end{tikzpicture}
\caption{English-label coverage and its complement among commercial
records. The complementary group includes any missing language labels.}
\label{fig:language}
\end{figure}

The complementary language partition contains 22,702 English-labeled
commercial records and 19,098 records with other or missing language labels
(Figure~\ref{fig:language}). The second group should not be interpreted as
19,098 validated non-English conversations: the audit does not separate
missing labels from specific languages. Similarly, a language label does
not establish language quality or monolingual content. A training selection
that requires both English and a minimum depth would need a joint query.
The available marginals alone cannot measure that selection.

\subsection{Contribution to the pooled commercial subset}
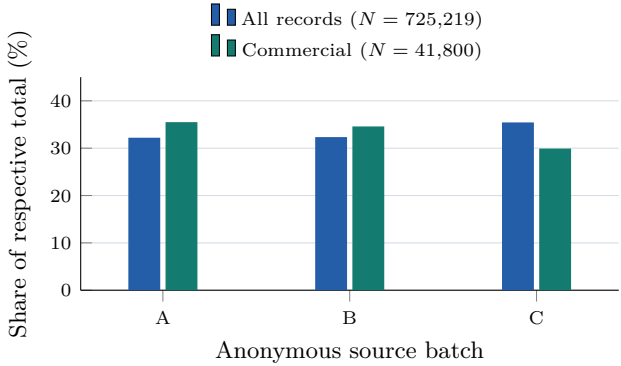
\begin{figure}[t]
\centering
\begin{tikzpicture}
\begin{axis}[width=\columnwidth,height=4.4cm,ybar,bar width=12pt,
 ymin=0,ymax=45,ytick={0,10,20,30,40},ylabel={Share of respective total (\%)},
 symbolic x coords={A,B,C},xtick=data,xlabel={Anonymous source batch},
 tick label style={font=\scriptsize},label style={font=\small},
 legend style={font=\scriptsize,at={(.5,1.03)},anchor=south,draw=none},
 ymajorgrids=true,grid style={StudyGrid},axis lines*=left,enlarge x limits=.22]
\addplot[fill=StudyBlue,draw=none] coordinates {(A,32.21)(B,32.34)(C,35.44)};
\addplot[fill=StudyTeal,draw=none] coordinates {(A,35.50)(B,34.59)(C,29.92)};
\legend{All records ($N=725{,}219$),Commercial ($N=41{,}800$)}
\end{axis}
\end{tikzpicture}
\caption{Source mixture before and after commercial-label selection.
Each series sums to 100\% before rounding.}
\label{fig:contribution}
\end{figure}

A batch's contribution to the commercial subset depends on both its size
and its commercial-label prevalence. Batch A supplies 32.21\% of the full
inventory but 35.50\% of commercial records; batch B supplies 32.34\%
and 34.59\%, respectively; batch C supplies 35.44\% and 29.92\%.
Figure~\ref{fig:contribution} compares these two denominators. Batch C is
the largest inventory source while contributing the smallest commercial
count. Consequently, selecting only commercially labeled examples changes
the source mixture even without deliberate source weighting.

For each batch, the ratio of commercial-subset share to inventory share
also equals its commercial prevalence divided by the pooled prevalence.
The ratios are approximately 1.10, 1.07, and 0.84 for A, B, and C. These
are descriptive composition ratios, not estimates of source quality. They
motivate preserving source membership in a training manifest and reporting
performance by held-out source rather than assuming pooled results transfer.

\section{Reference Architecture}
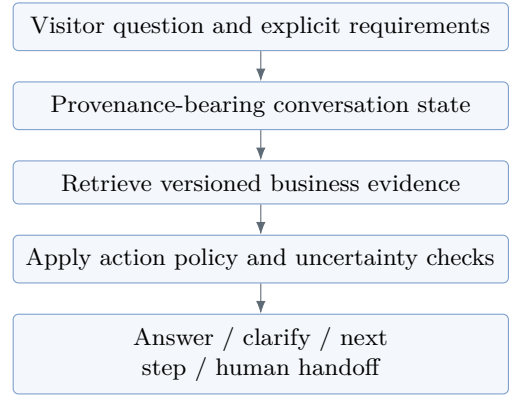
\begin{figure}[t]
\centering
\begin{tikzpicture}[node distance=4mm,
 box/.style={draw=StudyBlue!60,rounded corners=2pt,fill=StudyBlue!5,align=center,text width=6.2cm,font=\small,inner sep=5pt},
 arr/.style={-Latex,draw=StudyMuted}]
\node[box] (input) {Visitor question and explicit requirements};
\node[box,below=of input] (state) {Provenance-bearing conversation state};
\node[box,below=of state] (knowledge) {Retrieve versioned business evidence};
\node[box,below=of knowledge] (policy) {Apply action policy and uncertainty checks};
\node[box,below=of policy] (action) {Answer / clarify / next step / human handoff};
\draw[arr] (input)--(state);\draw[arr] (state)--(knowledge);
\draw[arr] (knowledge)--(policy);\draw[arr] (policy)--(action);
\end{tikzpicture}
\caption{Proposed agent architecture. This is a design specification,
not a verified implementation or trained-model evaluation.}
\label{fig:arch}
\end{figure}

The audit motivates explicit separation of commercial labels, useful training
examples, and live business actions. We propose a modular architecture with
four components: business knowledge, conversation state, response generation,
and constrained next-action selection. Figure~\ref{fig:arch} summarizes
the specification. No claim is made that the diagram describes an inspected
production implementation.

\subsection{Versioned business knowledge}
The knowledge layer contains offers, prices, eligibility, availability,
service areas, and approved explanations. Each item should retain its
source, validity period, and update version. Retrieval is conditioned on the
visitor's current question and accumulated explicit requirements. Conflicting
or stale evidence should trigger qualification or handoff, rather than an
unsupported answer. Business facts should be replaceable independently of
conversation-model weights.

\subsection{Provenance-bearing conversation state}
Let $u_t$ be the visitor message at step $t$, $s_{t-1}$ the previous state,
and $\Delta_t$ the explicit information extracted from $u_t$. A state update
is represented as
\begin{equation}
 s_t=\operatorname{Update}(s_{t-1},\Delta_t).
\end{equation}
The state stores needs, constraints, unanswered questions, and contact
preferences. Each value retains its supporting turn and whether it was
explicitly stated. Corrections supersede conflicting values; unknown fields
remain unknown. Inferred preferences should not silently become confirmed
facts. A short follow-up can rely on prior state without reproducing it.

\subsection{Constrained next actions}
The agent chooses among answering, clarifying, proposing a relevant next
step, and handing off. Given business evidence $K_t$ and policy $P$, its
allowed action set is
\begin{equation}
 \mathcal{A}_t=\operatorname{Allowed}(s_t,K_t,P).
\end{equation}
The response model selects within that set, rather than treating all fluent
continuations as permissible. A call request should match the visitor's
goal; contact collection should have a defined purpose; unavailable offers
and unsupported promises should be excluded. Regulated advice and requests
outside the configured scope should route to a human.

\subsection{Model adaptation}
Conversation-based adaptation can target clarification, appropriate evidence
use, constraint retention, and handoff. Suitable examples are not selected
solely because they have a commercial label. Context-response pairs require
rights clearance, privacy review, and validation of the response behavior.
Preference pairs should encode a defensible reason one response is better
than another. A model trained on such examples should be compared with its
base model under identical business knowledge, prompts, and tools.

\section{Collection and Quality Specification}
\subsection{From inventory to training eligibility}
An auditable pipeline distinguishes five objects: ingested records,
deduplicated records, commercially relevant records, cleared and
quality-accepted records, and records used in a named training job. A
large inventory cannot substitute for the last count. Messages, examples,
tokens, unique conversations, and repetitions across training epochs should
be reported separately.

For each accepted record, a restricted register should retain source,
collection date, permitted research and training uses, retention conditions,
and redistribution restrictions. Source-specific requirements take precedence
over a generic proprietary-corpus description. The present audit did not
verify this register, and does not certify training eligibility.

\subsection{Privacy and minimization}
Training and evaluation inputs should include only information necessary for
the intended task. Remove credentials, account identifiers, contact details,
and unnecessary personal or location attributes before example construction.
Pattern-based redaction alone may miss contextual identifiers, so selected
examples require contextual review. Transformation logs and raw data should
remain restricted. Public reporting should use aggregate statistics or
constructed examples; private transcript excerpts should not be included by
default. Neither an anonymous batch label nor a false safety flag establishes
that the underlying transcript is anonymized.

\subsection{Deduplication and split integrity}
Identifier uniqueness is the first check, not the final one. A training
pipeline should additionally compare normalized exact text and detect near
duplicates. Translations, repeated prompts, or successive snapshots of one
conversation can produce overlap even when hashes differ. Partitioning must
prevent those families from crossing training and evaluation splits.
When permitted participant keys are available, user-level separation provides
an additional control. Source-level holdouts test transfer rather than
memorization of one collection's style.

\subsection{Label and response audits}
Validate inherited commercial and transactional labels with stratified human
review. Sampling should include each source batch, language group, relevant
depth range, positive labels, and apparent negatives. Reviewers need an
operational rubric that separates product research, recommendation seeking,
action requests, and completed outcomes. Report disagreement, adjudication,
and a confusion matrix before treating the labels as ground truth.

Response quality needs a separate rubric: factual grounding, consistency
with explicit requirements, useful clarification, appropriate next action,
and handoff correctness. A commercially labeled dialogue may contain poor
assistant behavior. The uniform safety field observed here makes an
independent review particularly important; it is not a validated quality
certificate.

\subsection{Training manifests}
Each model version should preserve its base model, source versions, accepted
and excluded counts, deduplication rules, example construction, training
settings, split definitions, and output model identifier. This record is
the basis for a training-volume statement. The audited inventory contains
725,219 records, and the narrow commercial category contains 41,800;
these figures do not verify a claim that millions of commercial conversations
trained a Javo model. A larger separately documented training corpus may
exist, but it was not established by this audit.

\section{Evaluation Boundaries and Limitations}
The corpus is selected and heterogeneous. Unknown collection mechanisms and
label error limit generalization. Hash uniqueness does not exclude semantic
duplicates or participant dependence. Recorded depth is not role-validated.
The audit uses existing marginal metadata, so it cannot reconstruct buying
journeys, objection resolution, or outcomes.

Javo training is reported by the product owner and attributed to Javo's team.
No training manifest, checkpoint, deployed endpoint, retrieval trace, or
controlled evaluation was inspected. The architecture and quality pipeline
remain proposed specifications. The findings establish metadata composition,
not model superiority, training causality, or conversion uplift.

An evaluation of a trained model should freeze the model version and compare
it with the base model on leakage-controlled held-out cases. Blind human
review should calibrate automated ratings. Primary endpoints can include
factual accuracy, explicit-constraint satisfaction, useful clarification,
appropriate next-action selection, and correct handoff. Live lead quality,
booking, and revenue require separately linked outcomes and a controlled
comparison with consistent attribution. A simulated decision to buy should
not be reported as an actual sale.

\section{Conclusion}
The audited table contains 725,219 identifier-unique records, including
41,800 with a commercial label and 2,387 with a transactional label.
Commercial-label prevalence differs across source batches, and substantial
language and recorded-depth variation remains within the commercial category.
These results support explicit accounting of inventory, relevance, eligibility,
and training membership. The reference architecture separates business facts,
conversation state, and constrained actions; the quality specification makes
source rights, privacy, label validation, and split integrity inspectable.
The work provides a corpus audit and technical design rather than evidence
of sales improvement.

\section*{Data availability and reproducibility}
The analysis uses database-side aggregate queries and deterministic arithmetic.
The accompanying ancillary package includes anonymous aggregate counts,
the query specification, and a standard-library analysis script. It contains
no raw text, conversation hashes, participant identifiers, or internal source
names. Reproducing derived counts and shares from that snapshot is public;
repeating the live census requires authorized access to the restricted source.

\section*{Author and product disclosure}
Benjamin Tannenbaum is affiliated with Javo for this technical white paper.
Aiso provided access to the audited conversation data. Javo's team is
responsible for model training, as reported by the product owner. The product
context creates a commercial interest in the subject. This paper reports
no independently verified training count or model-performance result.

\end{document}